\documentclass[11pt,preprint,preprintnumbers,amsmath,amssymb,nofootinbib,aps]{revtex4}

\usepackage{graphics,color,array,dcolumn}
\usepackage{hyperref}
\usepackage{amsmath}
\usepackage{amssymb}
\usepackage{amsfonts}

\renewcommand\section{\@startsection {section}{1}{\z@}%
                                   {\normalfont\large\bfseries}}

\begin{document}

\preprint{ITP--UU--09/16}
\preprint{SPIN--09/16}

\title{\vspace*{4mm}{\bf Conformally reduced quantum gravity revisited\\[3ex]}}

\author{Pedro F. Machado$^a$}\email{p.f.machado@uu.nl}
\author{Roberto Percacci$^b$}\email{percacci@sissa.it}

\affiliation{{\footnotesize $^a\,$Institute for Theoretical Physics and Spinoza Institute\\
Utrecht University, 3508 TD, Utrecht, The Netherlands\\
$^b\,$ Department of Astronomy and Physics, University of Sussex,\\
and SISSA, via Beirut 4, I-34014 Trieste, Italy\\
and INFN, Sezione di Trieste, Italy\\}}

\begin{abstract}
\noindent 
Applying functional renormalization group methods, we describe
two inequivalent ways of defining the renormalization group of 
matter-coupled four dimensional gravity,
in the approximation where only the conformal factor is dynamical 
and taking the trace anomaly explicitly into account.
We make contact with earlier work and briefly discuss the presence
or absence of fixed points, depending on the truncation of the
action and other approximations.
\end{abstract}
\maketitle

\section{Introduction}

Although the conformal factor is not dynamical in classical general 
relativity, in quantum gravity its fluctuations could be as important, or even
more important, than those of the spin two components of the metric. 
It may thus be instructive to study a baby version of  quantum gravity where 
only the conformal part of the metric is allowed to fluctuate. 
We will refer to this theory as conformally reduced quantum gravity.
A popular approach to quantum gravity is to try and define 
the theory in the Euclidean. The following problem is then encountered:
when one uses the Hilbert action and fixes the sign in such a way
that spin two fluctuations have positive action, the conformal
fluctuations have negative action, and the Euclidean action
is unbounded from below \cite{ghp}.
This problem is often circumvented by an ad hoc rotation of the
integration contour in the complex plane;
a more satisfactory understanding of this issue, ultimately leading
to the same outcome, is based on a proper understanding of
the functional measure \cite{mm1,mottolameasure,dasguptaloll}.
At a less formal level, that is also reflected in the Causal 
Dynamical Triangulations approach \cite{ajl}.

On the other hand, while the Einstein action is the most important term at the classical level
and the obvious starting point for quantization, at the quantum level
other terms may also play a significant role.
At very high energies, higher derivative terms become important and - if they have the 
right signs - they can fix the problem of the unboundedness of the action.
At very low energies, nonlocal terms are expected to become relevant.
Among the latter, particularly interesting are those coming from
the Riegert action \cite{riegert}, which reproduces the conformal anomaly 
generated by matter loops.
The dynamics of conformally reduced gravity including such terms
has been studied in a series of papers by Antoniadis, Mazur and Mottola
\cite{am1,amm1,mazurmottola}. Following the logic of two dimensional 
conformal field theories, they argue that the theory has an 
infrared (IR) fixed point (FP), which could lead to screening of the cosmological 
constant and simulate dark energy  \cite{amm2,mottolavaulin}.

In a completely unrelated development, the renormalization group 
running of the gravitational couplings has been studied 
by use of a Functional Renormalization Group Equation \cite{reuter,dou}.
Again, the starting point for such applications has been the Einstein--Hilbert action
\cite{souma,reuterEH,fischerlitim},
but subsequently calculations have been extended to four-derivative
\cite{lauscherreuter,codellopercacci,Benedetti:2009rx}
or even higher terms \cite{largen,cpr1,machadosaueressig}
and some work also has been done on nonlocal terms \cite{machadosaueressig,rs}.
While the main aim of these calculations has been to establish the existence of
a FP with a finite number of UV-attractive directions, 
which could be used to define a sensible UV limit in a quantum field theory of gravity 
\footnote{For reviews of this asymptotic safety approach to quantum gravity,
see \cite{niedermaierreuter,niedermaierrev,revperc}.}, this type of analysis can be applied 
also to IR physics, and there have been works suggesting that FP behaviour
is responsible for astrophysical \cite{rw1}
and cosmological \cite{reutercosmology} effects.

The calculations of gravitational beta functions based on the FRGE have been carried
out mostly taking into account all the degrees of freedom of the metric
and truncating the action to a manageable number of terms. Conversely, there
have also been calculations where some degrees of freedom of
the metric were frozen, by requiring the existence of two Killing vectors \cite{niedermaier},
and infinitely many terms were kept in the action.
More recently, Reuter and Weyer have applied the FRGE to conformally reduced gravity 
\cite{reuterweyerI,reuterweyerII} and found, in certain truncations, a FP with very similar
properties as in the full theory.

The question then naturally arises, whether there exists a relation between
these FRGE beta functions and the beta functions computed by Antoniadis
and Mottola in \cite{am1}.
Establishing this relation is the one of the goals of the present work. 
Anticipating our results, we shall see that
Antoniadis and Mottola's beta functions {\it can} be obtained from the FRGE
within certain approximations, and applying a procedure that is different 
from Reuter and Weyer's.
We will explain and comment on this statement in detail in the following sections.

In the rest of this introduction, we will describe our approach
to the dynamics of the conformal factor, emphasizing possible alternatives. 
We will use the background field method and, following the procedure 
used both by Antoniadis et al. and Reuter et al, we will first fix a fiducial 
metric $\hat g_{\mu\nu}$
and consider only metrics which are conformally related to $\hat g_{\mu\nu}$:
\begin{equation}
\label{conformal}
g_{\mu\nu}=\phi^2\hat g_{\mu\nu}\ .
\end{equation}
The function $\phi$ is the conformal factor whose dynamics we wish to study.
Because it cannot vanish, we can choose it to be positive, and
in the following we will find it convenient to write $\phi=e^\sigma$.
The role of $\hat g_{\mu\nu}$ is simply to identify a conformal
equivalence class of metrics and to provide a reference point
in this equivalence class.
When restricted to the chosen conformal equivalence class,
the action, which originally is a functional of $g_{\mu\nu}$,
becomes a functional of $\hat g_{\mu\nu}$ and $\phi$, or equivalently of
$\hat g_{\mu\nu}$ and $\sigma$, which we will denote
\begin{equation}
\label{hatted}
\hat S(\hat g_{\mu\nu},\sigma)=S(e^{2\sigma}\hat g_{\mu\nu})=S(g_{\mu\nu})\ .
\end{equation}
No approximation is involved in this step.
Note that by construction $\hat S$ is invariant under the transformation
\begin{equation}
\label{weyl}
(\hat g_{\mu\nu},\sigma)\mapsto(e^{2\omega}\hat g_{\mu\nu},\sigma-\omega)\ ,
\end{equation}
for any function $\omega$.
We will refer to this as a Weyl transformation of $\hat g_{\mu\nu}$.
A priori, there is a slight risk of confusion between these transformations and
Weyl transformations of $g_{\mu\nu}$,
which are transformations $g_{\mu\nu}\mapsto e^{2\omega} g_{\mu\nu}$
\footnote{Note that if the original action is Weyl invariant,
in the sense that $S(e^{2\omega} g_{\mu\nu})=S(g_{\mu\nu})$,
then $\hat S$ is independent of $\sigma$.
}.
We will always try to make this difference clear.

We then apply the background field method to the conformal factor only.
In principle, there are different ways of doing this.
In \cite{reuterweyerI,reuterweyerII} the conformal factor is expanded as
\begin{equation}
\label{decompo}
\phi=\bar\phi+\delta\phi\ ,
\end{equation}
where $\bar\phi$ is the background.
Alternatively, one could write $\phi=e^{\sigma}$, $\bar\phi=e^{\bar\sigma}$
and expand
\begin{equation}
\label{bgexpand}
\sigma=\bar\sigma+\delta\sigma\ .
\end{equation}
Although these two procedures lead to similar results, they
are not strictly speaking equivalent within the approximations we will 
subsequently employ. In this paper we will follow the latter procedure,
as it is better adapted to the action of Weyl transformations.

When $\sigma$ is decomposed as in \eqref{bgexpand}, the transformation 
\eqref{weyl} can be attributed either to the fluctuation $\delta\sigma$
or to the background $\bar\sigma$. In the first case, we speak of ``quantum
Weyl transformations'', in the second, of ``background Weyl transformations''.
It is the latter transformations
\begin{equation}
\label{bgweyl}
(\hat g_{\mu\nu},\bar\sigma,\delta\sigma)
\mapsto(e^{2\omega}\hat g_{\mu\nu},\bar\sigma-\omega,\delta\sigma)\ 
\end{equation}
that one can preserve when using the background field method,
as we shall discuss in Section 4.
It is worth mentioning that this group does not play the role
of a gauge group, since it acts nontrivially 
on $\hat g_{\mu\nu}$, while in the conformal reduction
we only treat $\sigma$ as a quantum field.

\section{Dynamics of the conformal factor}

In this section we specify the class of gravitational actions we will study.
In order to avoid misunderstandings, let us stress from the outset that
these functionals will not be used as bare actions in the definition
of a functional integral, but rather as approximate forms for a
coarse grained quantum effective action.
With this proviso in mind, we will simply call these functionals ``actions''.
They will consist of one part which is local in the metric $g_{\mu\nu}$
and another part which can be seen as coming from the quantum loops of matter fields,
and which is nonlocal when written as a functional of $g_{\mu\nu}$.
Restricting ourselves to terms with at most four derivatives,
the local part is
\begin{equation}
\label{localaction}
S(g_{\mu\nu})=\int dx\,\sqrt{g}
\left[g_0+g_2 R+g_4 R^2\right]\ ,
\end{equation}
where $g_i$ are coupling constants of mass dimension $4-i$.
There are other terms one can write with four derivatives,
but they are either total derivatives (the Euler term, $\Box R$), 
or invariant under Weyl transformations of $g_{\mu\nu}$ 
(the Weyl tensor squared), and therefore independent of $\sigma$.
Using (\ref{conformal}) in (\ref{localaction}) and defining  $\hat\Box = \hat\nabla^2$, 
we have
\begin{equation}
\label{eff_action}
\hat S(\hat g_{\mu\nu},\sigma)
= \int dx\,\sqrt{\hat g} 
\Big[
g_0 e^{4\sigma} 
+g_2 e^{2\sigma}(\hat R-6\hat\Box\sigma-6(\hat\nabla\sigma)^2)
+g_4 (\hat R-6\hat\Box\sigma-6(\hat\nabla\sigma)^2)^2 
\Big]\ .
\end{equation}
In the following, we will need the linearized form of this expression.
Decomposing $\sigma$ as in (\ref{bgexpand}), 
and expanding to second order in $\delta\sigma$,
\begin{equation}
\begin{split}
\label{S_var}
\hat S^{(2)}
=&\int dx\,\sqrt{\hat g}\,
\delta\sigma
\Bigl[8 g_0 e^{4\bar\sigma}
+2 e^{2\bar\sigma}g_2\bigl(\hat R-6 \hat\Box\bar\sigma-6^2 (\hat\nabla\bar\sigma)^2 
-6\hat\nabla^\mu\bar\sigma\hat\nabla_\mu-3\hat\Box\bigr)
\\
&
+g_4\bigl(
-144\hat\Box\bar\sigma\hat\nabla^\mu\bar\sigma\hat\nabla_\mu
-144\hat\nabla^\mu(\hat\nabla\bar\sigma)^2\hat\nabla_\mu
+12\hat\nabla^\mu\hat R\hat\nabla_\mu
+144\hat R^{\mu\nu}\hat\nabla_\mu\bar\sigma\hat\nabla_\nu
\\
&
-72 (\hat\nabla\bar\sigma)^2\hat\Box
-144\hat\Box\bar\sigma\hat\Box
+12\hat R\hat\Box
+144(\hat\nabla^\mu\hat\nabla^\nu\bar\sigma
-\hat\nabla^\mu\bar\sigma\hat\nabla^\nu\bar\sigma)\hat\nabla_\mu\hat\nabla_\nu
\\
&
+36\hat\Box^2
\bigr)
\Bigr]
\delta\sigma\ .
\end{split}
\end{equation}

In addition, we will also consider the effect of minimally coupled massless matter. 
Introducing $n_S$ scalar fields $\phi$, $n_D$ Dirac fields $\psi$ 
and $n_M$ Maxwell fields $A_\mu$, the (gauge fixed) matter part of the action reads
\begin{equation}
S_{\rm mat}
=\int d^4 x \sqrt{g} \sum\left[
\tfrac{1}{2} \nabla_\mu \phi \nabla^\mu \phi 
+\bar{\psi}D \psi 
+\left(\tfrac{1}{4} F_{\mu\nu} F^{\mu\nu} +\tfrac{1}{2}(\nabla^\mu A_\mu)^2
-\bar c\,\Box c\right) \right]\,,
\end{equation}
where the sums extend over all particle species. Here, $D=\gamma^a e_a{}^\mu \nabla_\mu$,
is the Dirac operator ($e_a{}^\mu$ is the vierbein of $g_{\mu\nu}$)
and the last term above is the action for the ghost fields $\bar{c}, c$,
which arise when fixing the Lorentz gauge for the Maxwell fields.
Performing the conformal reduction \eqref{conformal} and applying the background field method
with the matter background fields set to zero,
the second variation of the matter and ghost parts of the action is then given by
\begin{equation}
\label{varmatter}
\begin{split}
S^{(2)}_{\rm mat} = &\int d^4 x \sqrt{\hat g}\sum
\Big[-\tfrac{1}{2}e^{2\bar\sigma}\delta\phi 
\left(\hat\Box+2\hat\nabla^\mu\bar\sigma\hat\nabla_\mu\right)\delta\phi
+e^{3\bar\sigma}\delta\bar{\psi}\left(\hat D+\gamma^a e_a{}^\mu\hat\Phi_\mu\right)\delta\psi \\
&\!\!\!\!\!\!\!\!-\tfrac{1}{2}\delta A_\nu\! \left(\hat g^{\mu\nu}(\hat\Box
+2\hat\nabla^\lambda\bar\sigma\hat\nabla_\lambda
+\hat\Box\bar\sigma+2(\hat\nabla\bar\sigma)^2)
-(R^{\mu\nu}\!-2\hat\nabla^\mu\hat\nabla^\nu\bar\sigma + 
2\hat\nabla^\mu\bar\sigma\hat\nabla^\nu\bar\sigma)\right)\delta A_\mu\\
&
+e^{2\bar\sigma}\delta\bar c\, 
\left(\hat\Box+2\hat\nabla^\mu\bar\sigma\hat\nabla_\mu\right)\delta c
\Big]
\,,
\end{split}
\end{equation}
where $\hat\Phi_\mu=2\hat e_{a\mu}\hat e_b{}^\nu\partial_\nu\sigma\Sigma^{ab}$.
These matter fields will contribute 
to the beta functions of the gravitational couplings $g_0$, $g_2$, $g_4$ \cite{dou,largen}. 
This is a purely local effect, which is related to the appearance of UV divergences
when the cutoff goes to infinity. On the other hand,
the presence of matter fields also gives rise to nonlocal terms,
among which there are those responsible for the conformal anomaly
\cite{duff}
\begin{equation}
\label{anomaly}
\langle T^\mu{}_\mu\rangle=
\frac{2}{\sqrt{g}}g_{\mu\nu}\frac{\delta\Gamma}{\delta g_{\mu\nu}}=
b\,C^2+b'E+\left(b''+\frac{2}{3}b\right)\Box R\ .
\end{equation}
Here,
$E=R_{\mu\nu\rho\sigma}R^{\mu\nu\rho\sigma}-4R_{\mu\nu}R^{\mu\nu}+R^2$
is the integrand of the Euler invariant,
$C^2=C_{\mu\nu\rho\sigma}C^{\mu\nu\rho\sigma}$
is the square of the Weyl tensor,
and the coefficients $b$ and $b'$ are
related to the number and species of matter fields and read
\begin{equation}
b=\frac{1}{120(4\pi)^2}\left(n_S+6n_D+12n_M\right)\ ,\qquad
b'=-\frac{1}{360(4\pi)^2}\left(n_S+11n_D+62n_M\right)\ .
\end{equation}

The last term in (\ref{anomaly}) can be obtained from the variation of a local
counterterm proportional to $\int dx\sqrt{g}R^2$, and so
the coefficient $b''$ is arbitrary.
This term is already accounted for in the local action \eqref{localaction}, and
it will be convenient to assume that $g_4$ has been redefined 
in such a way that $b''+\frac{2}{3}b=-\frac{2}{3}b'$.

The remaining two terms in the conformal anomaly (\ref{anomaly}) cannot be 
obtained as the variation of a local functional. Following \cite{am1}, those 
nonlocal counterterms responsible for generating this remaining part of the 
anomaly will also be 
taken into account. They constitute the Riegert action \cite{riegert} and are given by
\begin{equation}
\label{riegert}
W(g_{\mu\nu})=\frac{1}{8}\int dx\,\sqrt{g}
\left(E-\frac{2}{3}\Box R\right)
\Delta_4^{-1}
\left[2b\,C^2+b'\left(E-\frac{2}{3}\Box R\right)\right]\,,
\end{equation}
where $\Delta_4$ is the conformally covariant fourth order operator
\begin{equation}
\label{deltafour}
\Delta_4=\Box^2+2R^{\mu\nu}\nabla_\mu\nabla_\nu
-\frac{2}{3}R\Box+\frac{1}{3}\nabla^\mu R\nabla_\mu
\ .
\end{equation}

The defining property of this functional is that its variation
under an infinitesimal conformal transformation reproduces (\ref{anomaly}). 
One can also define a local functional
having the same property, at the expense of introducing an additional field.
This so called Wess--Zumino (WZ) action is (minus)
the change of the Riegert action under a finite conformal transformation,
\begin{equation}
\label{wz}
\Gamma_{WZ}(g_{\mu\nu},\sigma)=-W(e^{2\sigma}g_{\mu\nu})+W(g_{\mu\nu})\ .
\end{equation}
It is explicitly given by
\begin{equation}
\label{wzaction}
\Gamma_{WZ}(g_{\mu\nu},\sigma)
=-\int dx\,\sqrt{g}\left\{
b\,C^2\sigma+b'\left[\left(E-\frac{2}{3}\Box R\right)\sigma+2\sigma \Delta_4\sigma
\right]\right\}\ ,
\end{equation}
and, by construction, it satisfies the ``cocycle'' condition
(also called the Wess-Zumino consistency condition),
\begin{equation}
\label{cocycle}
\Gamma_{WZ}(e^{2\omega}g_{\mu\nu},\sigma-\omega)-\Gamma_{WZ}(g_{\mu\nu},\sigma)
+\Gamma_{WZ}(g_{\mu\nu},\omega)=0\ .
\end{equation}
Although $\sigma$ plays the role of a conformal
transformation in (\ref{wz}) , we can think of it as a new scalar field,
transforming under Weyl transformations as in (\ref{weyl}).
Then, by equation (\ref{cocycle}), the WZ action has the
same transformation as the nonlocal Riegert action (this property
motivates the sign in the definition of $\Gamma_{WZ}$).

Let us now treat the functional $W$ in the same way as the local action 
(\ref{localaction}).
As in (\ref{hatted}), we first define
$\hat W(\hat g_{\mu\nu},\sigma)=W(e^{2\sigma}\hat g_{\mu\nu})$, and
from equation (\ref{wz}) we then see that
\begin{equation}
\label{joy}
\hat W(\hat g_{\mu\nu},\sigma)=W(g_{\mu\nu})
=W(\hat g_{\mu\nu})-\Gamma_{WZ}(\hat g_{\mu\nu},\sigma)\ .
\end{equation}
Using equations (\ref{wz}) and (\ref{cocycle}), one can check
that this functional is indeed invariant under the Weyl transformations (\ref{weyl}).
Of course, if one is only interested in the dynamics of the conformal factor
for a fixed fiducial metric, the first term on the r.h.s. can be ignored,
but one should remember that it is essential for Weyl invariance.

Next, we introduce the background field decomposition (\ref{bgexpand}) for $\sigma$.
Defining the background metric $\bar g_{\mu\nu}=e^{2\bar\sigma}\hat g_{\mu\nu}$ and 
again using (\ref{cocycle}) and (\ref{wz}), we can write (\ref{joy}) as
\begin{equation}
\label{sorrow}
\hat W(\hat g_{\mu\nu},\sigma)
=W(\bar g_{\mu\nu})-\Gamma_{WZ}(\bar g_{\mu\nu},\delta\sigma)\ .
\end{equation}
Note that only the second term depends on the quantum field $\delta\sigma$.
From (\ref{wzaction}), we thus see that the expansion of $W$ to second order in the fluctuation 
is
\begin{equation}
\label{W_var}
\hat W^{(2)}=
2b'\int dx\,\sqrt{\hat g}\,
\delta\sigma{\hat\Delta}_4\delta\sigma=
2b'\int dx\,\sqrt{\bar g}\,
\delta\sigma\bar\Delta_4\delta\sigma\ ,
\end{equation}
where ${\hat\Delta}_4$ and $\bar\Delta_4$ are the operators (\ref{deltafour}) constructed 
with the metrics $\hat g_{\mu\nu}$ and $\bar g_{\mu\nu}$ respectively.

\section{The RG equation and the conformal anomaly}

In order to extract the beta functions of the theory, 
we make use of the Functional Renormalization Group Equation (FRGE) \cite{wetterich}
\begin{equation}
\label{frge}
\partial_t\Gamma_k=
\frac{1}{2}{\rm STr}\left(\frac{\delta^2\Gamma_k}{\delta\mathbf{\Phi}\delta\mathbf{\Phi}} 
+\mathcal{R}_k\right)^{-1}\partial_t\mathcal{R}_k\,,
\end{equation}
which describes the dependence of a coarse-grained effective action 
$\Gamma_k\left[\mathbf{\Phi}\right]$ on a momentum scale $k$. 
Here, $t:= \log{k/k_0}$, $\mathbf{\Phi}$ are all the fields present in the theory, 
STr is a functional (super)trace and $\mathcal{R}_k$ is an infrared cutoff
suppressing the contributions to the trace of eigenmodes with momenta below $k$.
The coarse grained effective action reduces to the ordinary effective action
in the limit $k\to 0$.

If one keeps all couplings on the r.h.s. fixed, including any couplings that
may appear in the definition of the cutoff $\mathcal{R}_k$, then one is effectively 
replacing the running effective action $\Gamma_k$ in the r.h.s. by a fixed
``bare'' action, and in this approximation the equation describes the running
of the one loop effective action in dependence of the cutoff $k$.
When applied to familiar quantum field theories in this approximation, the well known beta functions
are correctly reproduced.
But the FRGE is actually an exact equation
and it can be used to obtain nonperturbative results.
In particular, it has been applied to the calculation of beta functions 
for gravity in many different approximations,
always leading to the appearance of a nontrivial fixed point
\cite{souma,reuterEH,lauscherreuter,codellopercacci,Benedetti:2009rx,largen,cpr1,niedermaier}.

In the sequel, we will apply the FRGE to compute the beta functions
of conformally reduced gravity, in the spirit of the previous section.
This means that (in addition to the matter fields)
the only quantum field that we allow to fluctuate
is the conformal factor $\sigma$ (or equivalently $\phi$)
and the truncated running effective action is assumed to have the form
\begin{equation}
\label{gammak}
\Gamma_k(\hat g_{\mu\nu},\sigma,\mathbf{\psi})=
\hat S(\hat g_{\mu\nu},\sigma)
+\hat W(\hat g_{\mu\nu},\sigma)
+\hat S_{\rm mat}(\hat g_{\mu\nu},\sigma,\mathbf{ \psi}) \equiv
\Gamma_k^{\rm grav}(\hat g_{\mu\nu},\sigma)
+\hat S_{\rm mat}(\hat g_{\mu\nu},\sigma,\mathbf{ \psi})
\end{equation}
where $\hat S$, $\hat S_{\rm mat}$ and $\hat W$ are given by equations 
(\ref{eff_action},\ref{varmatter},\ref{joy}), 
and $\mathbf{ \psi}$ collectively denotes all matter fields.
However, not all terms will run. $\hat S_{\rm mat}$ 
does not change because the fields have no selfinteractions
and $\hat W$ does not change because its coefficients $b$ and $b'$ 
are fixed functions of the number of matter fields. Thus, only the RG flow
of the couplings $g_0$, $g_2$ and $g_4$ will be calculated, while 
$\hat W$ and $\hat S_{\rm mat}$ will be kept fixed.

Although only $\mathbf{ \psi}$ and $\sigma$ fluctuate, the action still
depends parametrically upon the fiducial metric $\hat g_{\mu\nu}$ and,
as long as the Weyl invariance (\ref{weyl}) is preserved,
the running effective action $\Gamma_k$ can be regarded as
a functional of a single metric $g_{\mu\nu}$.
As discussed in \cite{floreanini}, in quantizing the theory of the conformal factor 
described by some action $\hat S(\hat g,\sigma)$ one faces a choice: 
the cutoff can be constructed with the fiducial metric $\hat g$ or with the 
background metric $\bar g_{\mu\nu}$. 
The former choice breaks the invariance (\ref{weyl}), because it introduces a dependence on 
$\hat g$ which is not accompanied by a corresponding dependence on $\sigma$. 
The latter choice instead respects the invariance.
For this reason, we shall call these two procedures the 
``Weyl--breaking'' 
and the 
``Weyl--invariant''
procedure respectively (and we emphasize here that we refer
to the Weyl transformations of the metric $\hat g_{\mu\nu}$, not of the metric $g_{\mu\nu}$).

These considerations apply both to UV and IR cutoffs.
A UV cutoff can be regarded as part of the definition of the functional integral.
In this context, the ``Weyl--breaking'' procedure corresponds to using
the translation invariant measure, 
while the ``Weyl--invariant'' procedure corresponds to using the
Weyl--invariant measure 
\footnote{See \cite{mottolameasure,bbm} for a discussion of these integration measures.},
and similar considerations also apply to the integration measures over the matter fields.
In the approach based on the FRGE, the beta functions give the
dependence of the {\it renormalized} couplings on the coarse
graining scale $k$, and these UV issues are completely immaterial.
Even though the FRGE is formally derived from a functional integral
which would require a UV regulator to be defined, the trace on the r.h.s.
of (\ref{frge}) is automatically UV convergent due to the properties of
the IR cutoff ${\cal R}_k$. Therefore, there is no need to specify any UV regulator.
In the following sections, when we talk about Weyl--invariant and Weyl--breaking
procedures, we then refer to the construction of the IR cutoff  ${\cal R}_k$,
which is used to define the coarse graining of the effective action.

Still, to avoid possible misunderstandings, it is useful to
comment here on the significance of the anomaly in the context of the FRGE.
The conformal anomaly arises when the ``classical'' bare action 
is Weyl invariant but the measure is not, and hence neither is
the quantum effective action.
This is true also for the coarse grained affective action $\Gamma_k$,
for any value of the coarse graining (IR cutoff) scale $k$.
In an ``anomalous'' theory, the running effective action
will thus be noninvariant even in the limit $k\to\infty$, if the limit exists.
Now, one could take the point of view that the functional integral and the
bare action are merely formal constructions devoid of physical content,
and that all the physics is contained in the running effective action $\Gamma_k$.
One would then never see an ``anomaly'': one simply has a quantum theory where
Weyl invariance is broken at all scales.
The ``anomaly'' would only be seen if one tried to reconstruct the
``classical'' (bare) action that corresponds to the given effective action 
(see \cite{manriquereuter} for a general discussion of this reconstruction
problem and \cite{reuterwetterich} for a specific discussion of
functional measures in the context of a FRGE--based treatment
of two dimensional Liouville theory).
Still, while this may be instructive and even useful for some purposes,
one would not learn anything new about the physics considered here
by doing this.

This discussion provides an answer to a question that may arise in this context.
The term $\hat W$ is usually regarded as (part of) the effective action
obtained by integrating out the matter fields, and one may wonder
why we keep $\hat S_{\rm mat}$ and $\hat W$ simultaneously in the action.
The reason for this is that we apply the same coarse graining scale to
the gravitational degree of freedom $\sigma$ and to the matter fields $\psi$.
So, as we do not first completely integrate out the matter fields,
$\hat S_{\rm mat}$ must still be present in the action
\footnote{Of course, at a given energy scale $k$ the degrees of freedom with masses $m>k$
will decouple and therefore in practice we need to consider only the
degrees of freedom with masses $m<k$. In the IR limit, only massless fields matter.}.
On the other hand,
as the term $\hat W$ describes the effect of the conformal anomaly, 
it is also present for any finite value of the coarse graining scale.
(In any case, one can easily remove from the beta functions the terms coming from
$\hat S_{\rm mat}$ and/or $\hat W$ if one so wishes.)

From here on, let us assume that the functional measure of the matter fields
in the functional integral is not Weyl invariant, so that 
$\Gamma_k$ contains the term $\hat W$.
The invariance, or lack thereof, of the functional measure of $\sigma$ 
only affects the numerical value of the coefficients in $\hat W$
\cite{am1,reuterwetterich}, and we do not need to commit ourselves
to a particular choice for our calculations in the next sections.
We will not discuss here the possibility of recovering Weyl invariance
in the limit $k\to 0$. This has been discussed in the two dimensional
case in \cite{reuterwetterich} and similar considerations could also be
applied in four dimensions.
We will focus instead on the form of the beta functions.

In \cite{reuterweyerI}, it has been explained in detail that choosing
the IR cutoff in a  Weyl--invariant way corresponds to implementing 
background independence in the quantum theory. 
This is the procedure that is always followed in the FRGE approach to 
asymptotic safety, also when the full metric is dynamical.
In the next two sections, we will compare the results of using the
Weyl--invariant and the Weyl--breaking implementations of the IR cutoff.

\section{The Weyl--invariant procedure}

In \cite{reuterweyerI}, the beta functions of the conformal reduction of gravity
with the Hilbert action were computed using a ``background independent'' IR cutoff,
constructed from the background metric $\bar g_{\mu\nu}$.
In this section, we follow a similar procedure, but rather than applying
the background field method to $\phi$, viz. \eqref{decompo}, we apply it 
to $\sigma$, viz. \eqref{bgexpand}, as we find that  the 
behavior of the theory under Weyl transformations is easier to understand
in this way. 
We also extend the results by including the 
effect of the $R^2$ term and of the Riegert action, which will be
needed when comparing with the beta functions of \cite{am1}, as well as the effect of the local matter
contribution.

The FRGE \eqref{frge} requires the second variation
$\frac{\delta^2 \Gamma_k^{\rm grav}}{\delta\sigma\delta\sigma}$,
which can be immediately read off equations \eqref{S_var} and \eqref{W_var}.
Those variations are written in terms of operators constructed with
the fiducial metric $\hat g_{\mu\nu}$ and the background field $\bar\sigma$, but,
in order to guarantee that (background) Weyl invariance is preserved,
it is convenient to rewrite them in terms of the metric $\bar g_{\mu\nu}$.
For the Riegert action, this has already been done in \eqref{W_var}.
For the rest, we observe that, since \eqref{eff_action} is invariant
under Weyl tranformations and $\delta\sigma$ is invariant under
background Weyl transformations \eqref{bgweyl}, the operator
appearing in square brackets in \eqref{S_var} must also be invariant
under background Weyl transformations.
Indeed, this can be verified by a straightforward if somewhat tedious calculation.
We can then apply a transformation \eqref{bgweyl} with parameter
$\omega=\bar\sigma$ to the second variation, leading to the substitutions 
$\hat g_{\mu\nu}\to\bar g_{\mu\nu}$ and $\bar\sigma\to 0$ in \eqref{S_var},
so that
\begin{equation}
\label{sym_var_inv}
\frac{\delta^2 \Gamma_k^{\rm grav}}{\delta\sigma\delta\sigma} =
\sqrt{\bar g}\Bigl[ 16 g_0 + 4 g_2(\bar R - 3\bar\Box)+
g_4(72 \bar\Box^2 + 24\bar R \bar\Box + 24\bar\nabla^\mu \bar R \bar\nabla_\mu)
+4b'\bar\Delta_4 \Bigr]\,.
\end{equation}
For our purposes, it will be enough to consider the case when $\hat g_{\mu\nu}$ 
is a space of constant curvature, for which
\begin{equation}
\frac{\delta^2 \Gamma_k^{\rm grav}}{\delta\sigma\delta\sigma} = 
\sqrt{\bar g}
\left[16 g_0+4g_2\bar R
+\left(\left(24g_4-\tfrac{2}{3}b'\right)\bar R-12g_2\right)\bar\Box
+\left(72g_4+4b'\right)\bar\Box^2\right]
\,.
\end{equation}
A similar reasoning applies to the second variation of the local matter contribution. 

There is a vast freedom in defining a cutoff, and one choice that presents
itself is that of an operator whose eigenfunctions are taken as a basis
in the functional space that one is integrating over.
The cutoff is then imposed on the eigenvalues of this operator
\footnote{To avoid possible misunderstandings, let us stress that the functional
trace in \eqref{frge} is obviously independent of any choice of functional
basis. What we are saying here is that putting a cutoff on the eigenvalues
of different operators leads effectively to different cutoff procedures.}.

We begin by following \cite{reuterweyerI} and choose this operator to
be $-\bar\Box$. As in \cite{cpr1} we will call this a ``type I cutoff''.
We then choose $\mathcal{R}_k$ such that it leads to the
replacement of $-\bar\Box$ by $P_k(-\bar\Box)=-\bar\Box+R_k(-\bar\Box)$
in the inverse propagator,
where $R_k$ is a suitable profile function suppressing the propagation 
of field modes below the scale $k$. 
In our subsequent calculations, we will chose as this function the so-called 
optimized cutoff \cite{optimized} $R_k(z)=(k^p-z)\Theta(k^p-z)$, where 
$\Theta$ is the step function and $p$ is the order of the operator $z$. 
Following this prescription leads to
\begin{equation}
\mathcal{R}_k= 
\sqrt{\bar g}\Big[
(72g_4+4b')(P_k^2-\bar\Box^2)
-\left(\left(24g_4-\tfrac{2}{3}b'\right)\bar R-12g_2\right)R_k\Big]
\,,
\end{equation}
and we thus arrive at 
\begin{equation}
\begin{split}
\partial_t \Gamma_k=&\frac{1}{2}{\rm Tr}\,
\Biggl\{
\frac{
\left[6g_2-\left(12g_4-\tfrac{1}{3}b'\right)\bar R+4(18g_4+b')P_k\right]\partial_t R_k}
{8g_0+2g_2-\left(\left(12g_4-\tfrac{1}{3}b'\right)\bar R-6g_2\right)P_k+(36g_4+2b')P_k^2}
\\
&
\quad +\frac{6\beta_2R_k+36\beta_4\left(P_k^2-\bar\Box^2-\frac{\bar R}{3}R_k\right)}
{8g_0+2g_2-\left(\left(12g_4-\tfrac{1}{3}b'\right)\bar R-6g_2\right)P_k+(36g_4+2b')P_k^2}
\Biggr\}
\\
&
\quad +\frac{n_S}{2}{\rm Tr}\frac{\partial_t R_k}{P_k}
-\frac{n_D}{2}{\rm Tr}\frac{\partial_t R_k}{P_k + \tfrac{\bar R}{4}} 
+ \frac{n_M}{2}{\rm Tr}\frac{\partial_t R_k}{P_k + \tfrac{\bar R}{4}}
-n_M{\rm Tr}\frac{\partial_t R_k}{P_k}
\ ,
\end{split}
\end{equation}
where we have defined $\beta_i =\partial_t g_i$, and the terms containing $\beta_i$
come from deriving the couplings that are contained in ${\cal R}_k$.
Note that all dependence on $\hat R$ and $\sigma$ is through
the background metric $\bar g_{\mu\nu}$, which is inert
under the background Weyl transformations \eqref{bgweyl}.
As the quantum field is also inert,
background Weyl invariance is respected.
Consequently, the flow will preserve the form of the action \eqref{eff_action},
and to extract the beta functions of $g_2$ and $g_4$ we can 
isolate the coefficient of any one of the operators that they multiply.
We evaluate the functional trace on the right-hand side of the FRGE using
the heat kernel expansion of the operator $-\bar\Box$
(using methods explained in, e.g., Appendix A of \cite{cpr1})
and then equate the coefficient of $\bar R^i$ with $\beta_i$.
This gives
\begin{equation}\label{typeI_winv}
\begin{split}
\beta_0 = & c_0 k^4
+\frac{(6g_2+(72g_4+4b')k^2+\beta_2+9\beta_4k^2)k^6}
{64\pi^2\left(4g_0+3g_2k^2+(18g_4+b')k^4\right)} \ ,\\
\beta_2=&c_2k^2
+\frac{
12\left(2g_0(g_2+6g_4k^2)+9g_4(g_2+18g_4k^2)k^4\right)k^4+2(10g_0+3g_2k^2+81g_4k^4)b'k^6+3b^{\prime 
2}k^{10}}
{192\pi^2\left(4g_0+3g_2k^2+(18g_4+b')k^4\right)^2}\\
&\quad
+\beta_2\frac{12g_0+3g_2k^2+(90g_4+2b')k^4}{384\pi^2\left(4g_0+3g_2k^2+(18g_4+b')k^4\right)^2}
+\beta_4\frac{(16g_0-6g_2k^2+(180g_4+b')k^4)k^2}{128\pi^2\left(4g_0+3g_2k^2+(18g_4+b')k^4\right)^2}\ ,\\
\beta_4 = & \,c_4
+\Big\{
3\big(g_0^2(464g_2-5952g_4k^2)k^2-24g_0(31g_2^2+198g_2g_4k^2+72g_4^2k^4)k^4\\
&\qquad
+9(29g_2^3-744g_2^2g_4k^2-12060g_2g_4^2k^4+64368g_4^3k^6)k^6\big) \\
&\qquad
+4\big(472g_0^2-198g_0(g_2-32g_4k^2)k^2
+9(14g_2^2-465g_2g_4k^2+3186g_4^2k^4)k^4\big)b'k^4\\
&\qquad
+\left(344g_0+75g_2k^2+1512g_4k^4\right)b^{\prime 2}k^8+28b^{\prime 3}k^{12}
\Big\}
\Big/34560\pi^2\left(4g_0+3g_2k^2+(18g_4+b')k^4\right)^3\\
& 
+\beta_2k^2\Big\{
464g_0^2+24g_0(g_2-354g_4k^2)k^2+27(3g_2^2-44g_2g_4k^2+1548g_4^2k^4)k^4\\
&\qquad
+8(14g_0+3(g_2+36g_4k^2)k^2)b'k^4+9b^{\prime 2}k^8\Big\}
\Big/23040\pi^2\left(4g_0+3g_2k^2+(18g_4+b')k^4\right)^3\\
& 
-\beta_4k^4\Big\{
496g_0^2+72g_0(17g_2+22g_4k^2)k^2+99g_2^2k^4+9828g_2g_4k^6-22356g_4^2k^8\\
&\qquad
+2(164g_0+93g_2k^2+918g_4k^4)b'k^4+36b^{\prime 2}k^8\Big\}
\Big/3840\pi^2\left(4g_0+3g_2k^2+(18g_4+b')k^4\right)^3\ ,\\
\end{split}
\end{equation}
where the constants $c_i$ are the local contribution of matter:
\begin{equation}\label{matcont}
\begin{split}
c_0=&\frac{1}{32\pi^2}\left(n_S-4n_D+2n_M\right)\ ,\\
c_2=&\frac{1}{96\pi^2}\left(n_S+2n_D-4n_M\right)\ ,\\
c_4=&\frac{1}{34560\pi^2}\left(29n_S-11n_D-62n_M\right)\ .
\end{split}
\end{equation}
The above formulae should be looked upon as a system
of linear equations for the beta functions $\beta_i$.
The beta functions themselves are obtained by solving these equations
and are somewhat complicated rational functions of the couplings.
If one deletes all the terms containing $\beta_i$ in the r.h.s.,
the remaining terms are the beta functions in the one loop approximation.

It is instructive to rederive the beta functions using a different cutoff procedure.
Instead of using the operator $-\bar\Box$ as defining the basis
in function space, we can use the fourth order operator
\begin{equation}
\bar{\cal O}\equiv \frac{1}{\sqrt{\bar g}(72 g_{44}+4b')}\frac{\delta^2 
\Gamma_k^{\rm grav}}{\delta\sigma\delta\sigma}
=\bar\Box^2+\ldots\,.
\end{equation}
Then, we define the cutoff
$\mathcal{R}_k=\sqrt{\bar g}(72 g_{44}+4b')R_k(\bar{\cal O})$,
where $R_k$ is the function defined above,
such that it leads to the replacement of 
$\frac{\delta^2 \Gamma_k^{\rm grav}}{\delta\sigma\delta\sigma}$ by 
$\sqrt{\bar g}(72 g_{44}+4b')P_k(\bar{\cal O})=\sqrt{\bar g}(72 g_{44}+4b')(\bar{\cal O}
+R_k(\bar{\cal O}))$.
This is called a ``type III cutoff''.
In this case, the FRGE simply reduces to
\begin{equation}
\label{simple}
\partial_t \Gamma_k = \frac{1}{2}{\rm Tr}
\frac{\partial_t R_k(\bar{\cal O})}{P_k(\bar{\cal O})} 
+\frac{n_S}{2}{\rm Tr}\frac{\partial_t R_k}{P_k}
-\frac{n_D}{2}{\rm Tr}\frac{\partial_t R_k}{P_k + \tfrac{\bar R}{4}} 
+ \frac{n_M}{2}{\rm Tr}\frac{\partial_t R_k}{P_k + \tfrac{\bar R}{4}}
-n_M{\rm Tr}\frac{\partial_t R_k}{P_k}\ .
\end{equation}
where the argument of the functions $R_k$ and $P_k$ in the
matter traces is still $-\bar\Box$.

Restricting ourselves to the one loop approximation, we arrive at the beta functions
\begin{equation}
\begin{split}
\beta_0 =&\,c_0k^4 + \frac{1}{32\pi^2}\left(
\frac{ 9 g_2^2 - 8 g_0 (b'+18 g_4)}{(b' + 18 g_4)^2} - \frac{6 g_2 k^2}{
 b' + 18 g_4} + 4 k^4
 \right)\,,\\
%
\beta_2 = &\, c_2 k^2 + \frac{1}{32\pi^2}\left( \frac{-2 b' g_2 -90 g_2 g_4 + 30 b' g_4 k^2 + 432 g_4^2 
k^2}{(b' + 18 g_4)^2}\right)\,,\\
\beta_4 =&\, c_4+ \frac{1}{32\pi^2}\left(\frac{29}{540} 
- \frac{b'^2 + 36 b' g_4 - 2592 g_4^2}{
 36 (b' + 18 g_4)^2}\right)\,.\\
\end{split}
\end{equation}

We can compare these beta functions with the corresponding
type I counterparts in the one loop approximation,
i.e. dropping the terms that contain $\beta_2$ or $\beta_4$ on the r.h.s. of \eqref{typeI_winv}.
The differences that one observes are a manifestation of the scheme dependence
of the results.
We expect only the one loop part of $\beta_4$, in the limit $k^2 \gg g_2$, $k^4 \gg g_0$,
 to be scheme--independent.
To this effect, one should expand the denominators of the type I
beta functions in powers of $g_0$ and $g_2$ and compare term by term.
Then one sees that the leading term of the expansion of $\beta_4$ is equal to
\begin{equation}
\label{universal}
\frac{7 b^{\prime2}+252 g_4 b'+24138 g_4^2}{8640 \pi ^2 \left(b'+18 g_4\right)^2}
\end{equation}
with both cutoff types, as expected.
This is then a really ``universal'' result.
Higher order terms of $\beta_4$ and all the terms in $\beta_0$ and $\beta_2$
are scheme--dependent. This does not make them physically unimportant,
although extracting physical predictions from them requires more work and more care.

One result from the scheme--dependent terms that should be scheme--independent
is the existence of a fixed point.
A fixed point is a simultaneous zero for the beta functions of 
the dimensionless variables $\tilde g_i = k^{-d_i} g_i$
(with $d_0=4$, $d_2=2$ and $d_4=0$), which are given by
\begin{equation}
\label{dimensionlessbeta}
\partial_t\tilde g_i=-d_i\tilde g_i+k^{-d_i}\beta_i\,.
\end{equation}
It is noteworthy that when the beta functions are written out
in terms of the variables $\tilde g_i$, the cutoff scale $k$ does not appear
explicitly anymore, in accordance with the general expectation that
the flow equations are autonomous.

We will now briefly discuss the fixed points of \eqref{typeI_winv}.
To make contact with \cite{reuterweyerI}, we begin by considering 
the case when matter is absent.
Further reducing ourselves to the Einstein--Hilbert truncation, where $g_4=0$,
the above equations admit a fixed point at $\tilde g_0=0.00404$ and
$\tilde g_2=-0.007296$, which correspond to 
$\tilde\Lambda=0.277$ and $\tilde G=2.727$.
These values are numerically very close to the result of \cite{reuterweyerI};
the residual discrepancy can be attributed to the fact that
we take $\sigma$ as the quantum field whereas \cite{reuterweyerI} use $\phi$,
and that imposing a cutoff on fluctuations of $\sigma$ is different from
imposing a cutoff on fluctuation of $\phi$.

Let us now extend the truncation to include the $R^2$ term.
If we set $\tilde g_0=\tilde g_2=0$, $\beta_4$ reduces to the
``universal'' expression \eqref{universal}, and in the absence of matter $b'=0$,
which leads to $\beta_4=149/17280\pi^2$.
It is not conceivable that higher order terms exactly cancel this term,
so this indicates that $1/g_4$ is asymptotically free,
and there is no FP for $\tilde g_0=\tilde g_2=g_4=b'=0$.
A more detailed analysis shows that equations \eqref{typeI_winv}
do not admit any nontrivial fixed point with positive $G$. 
\footnote{This is also the case when one uses the parametrization \eqref{decompo}.}

The fixed point may reappear when higher powers of curvature are allowed. 
In fact, it has been observed in \cite{cpr1,machadosaueressig} that the results
of the $R^2$ truncation are somewhat atypical and 
change significantly when higher order couplings are taken into account.
In any case, the fixed point does reappear when one takes
matter field contributions into account.
In the case of, e.g., one massless Maxwell field 
and no massless Dirac and scalar fields, we find a fixed point at 
${\tilde g_0 = 0.00135,\,\tilde g_2 = -0.00168,\, g_4 = 0.00036}$, 
corresponding to $\tilde\Lambda=0.401$ and $\tilde G=11.83$.
That the FP of pure gravity is quite close to the boundary of the existence
region in $n_S$-$n_D$-$n_M$ space has been also observed in \cite{perini}.

\section{The Weyl-breaking procedure}

We now want to calculate the beta functions of conformally reduced gravity
when the cutoff is defined by means of the fiducial metric $\hat g_{\mu\nu}$,
instead of the background $\bar g_{\mu\nu}$.
We will first use a type I cutoff.
To this effect, we follow the same steps as in the previous section, with the 
crucial difference that the IR cutoff is imposed on the spectrum of $-\hat\Box$, 
rather than $-\bar\Box$.
This introduces a dependence on $\hat\Box$ which is not compensated by the presence of
$e^{\bar\sigma}$ factors, and therefore breaks Weyl invariance.
As a consequence, the special form of the action \eqref{eff_action} will no longer
be preserved by the flow.
To see this, it is instructive to consider the slightly more general class of actions
\begin{equation}
\begin{split}
\label{gen_eff_action}
\hat S(\hat g_{\mu\nu},\sigma)=
&\!\int d^4x\,\sqrt{\hat g}
\Big[
g_0 e^{4\sigma}\!
+e^{2\sigma}(g_{21}\hat R
-6g_{22}\hat\Box\sigma
-6g_{23}(\hat\nabla\sigma)^2)
+g_{41}\hat R^2
-12g_{42}\hat R\hat\Box\sigma\\
&-12g_{43}\hat R(\hat\nabla\sigma)^2
+36g_{44}(\hat\Box\sigma)^2
+36g_{45}((\hat\nabla\sigma)^2)^2
+72g_{46}\hat\Box\sigma(\hat\nabla\sigma)^2
\Big]\ ,
\end{split}
\end{equation}
which are invariant under (global) scale transformations. These actions
become invariant under (local) Weyl transformations when the couplings $g_{2i}$ ($i=1,2,3$)
and $g_{4j}$ ($j=1,\ldots,6$) are separately equal. If the flow preserved
local Weyl--invariance, the beta functions of the $g_{2i}$ and $g_{4j}$
should then also be the same. We will shortly show that this is not the case.

For the sake of comparison with the preceding section, we
begin by analyzing the situation when the background $\bar\sigma$ is constant,
which allows us to extract the beta equations for the couplings $g_0,g_{21}$ and
$g_{41}$. In this case,
\begin{equation}
\begin{split}
\frac{\delta^2 \Gamma_k^{\rm grav}}{\delta\sigma\delta\sigma} = \sqrt{\hat g}&
\Bigl[16g_0e^{4\bar\sigma}+4g_{21}e^{2\bar\sigma}\hat R
+\left((24g_{43}-\tiny\frac{2}{3}b')\hat R-12(2g_{22}-g_{23})e^{2\bar\sigma}\right)\hat\Box\\
&+\left(72g_{44}+4b'\right)\hat\Box^2\Bigr]\ .
\end{split}
\end{equation}
Choosing the cutoff $\mathcal{R}_k$ such that $-\hat\Box$ is replaced by 
$P_k(-\hat\Box)=-\hat\Box+R_k(-\hat\Box)$ in the inverse propagator then leads to
\begin{equation}
\mathcal{R}_k= \sqrt{\hat g}
\Big[
-\left((24g_{43}-\tiny\frac{2}{3}b')\hat R-12(2g_{22}-g_{23})e^{2\bar\sigma}\right)R_k
+\left(72g_{44}+4b'\right)(P_k^2-\hat\Box^2)
\Big]\ .
\end{equation}
The cutoff for the matter fields follows the same logic.
For example, the inverse propagator of the scalar field is
$-e^{2\bar\sigma}\hat\Box$ and we choose the cutoff
$e^{2\bar\sigma}R_k(-\hat\Box)$, such that the modified
inverse propagator is $e^{2\bar\sigma}P_k(-\hat\Box)$.
Note that, in this way, the exponentials cancel between numerator
and denominator in the FRGE, and the matter contribution
is $\bar\sigma$--independent.
The FRGE thus reads
\begin{equation}
\begin{split}
\label{breaking}
\partial_t \Gamma_k=&\frac{1}{2}{\rm Tr}\,
\Biggl\{
\frac{\left[6(2g_{22}-g_{23})e^{2\bar\sigma}
-\left(12g_{43}-\tfrac{1}{3}b'\right)\bar R
+4(18g_{44}+b')P_k\right]\partial_t R_k}
{8g_0e^{4\bar\sigma}\!+2g_{21}e^{2\bar\sigma}\hat R
+\!\left(6e^{2\bar\sigma}(2g_{22}-g_{23})
\!-\!\left(12g_{43}-\tfrac{1}{3}b'\right)\!\hat R\right)\!P_k
+(36g_{44}+2b')P_k^2}
\\
&
+\frac{6(2\beta_{22}-\beta_{23})e^{2\bar\sigma}R_k
+36\beta_{44}(P_k^2-\hat\Box^2)-12\beta_{43}\hat R R_k}
{8g_0e^{4\bar\sigma}\!+2g_{21}e^{2\bar\sigma}\hat R
+\!\left(6e^{2\bar\sigma}(2g_{22}-g_{23})
\!-\!\left(12g_{43}-\tfrac{1}{3}b'\right)\!\hat R\right)\!P_k
+(36g_{44}+2b')P_k^2}
\Biggr\}
\\
&
\quad +\frac{n_S}{2}{\rm Tr}\frac{\partial_t R_k}{P_k}
-\frac{n_D}{2}{\rm Tr}\frac{\partial_t R_k}{P_k + \tfrac{\hat R}{4}} 
+ \frac{n_M}{2}{\rm Tr}\frac{\partial_t R_k}{P_k + \tfrac{\hat R}{4}}
-n_M{\rm Tr}\frac{\partial_t R_k}{P_k}\ .
\end{split}
\end{equation}
Evaluating the trace via a heat kernel expansion of $-\hat\Box$ and reading off the
coefficients of $e^{4\bar\sigma}$, $e^{2\bar\sigma}\hat R$ and $\hat R^2$, we then
arrive at the beta functions 
\begin{equation}
\begin{split}
\label{betaI}
\beta_0 =&\, 
\frac{9\left((g_{23}-2g_{22})^2-16g_0g_{44}\right)-8b'g_0}{32\pi^2(18g_{44}+b')^2}
+\frac{3 (g_{23} -2 g_{22})}{32\pi^2(18g_{44}+b')^2}\beta_{22}\\
&+\frac{3\left(2 g_{22} - g_{23}\right)}{64\pi^2(18g_{44}+b')^2}\beta_{23}
-\frac{9(4b'g_0-9((g_{23}-2g_{22})^2-8g_0g_{44}))}{64\pi^2(18g_{44}+b')^3}
\beta_{44}\,,\\
\beta_{21} =&\frac{9((g_{23}-2 g_{22})(g_{44}+2 g_{43}) -2 g_{21}g_{44})-b' g_{21}}
{16\pi^2(18 g_{44} + b')^2}+\frac{b'+27 g_{44} + 18 g_{43}}{192 \pi^2 (18 g_{44} + b')^2}
(2\beta_{22}-\beta_{23})\\
&\,-\frac{3 (b' (3 g_{21} + 2 g_{22} - g_{23}) +18 (3 g_{21} g_{44} + 2 (2 g_{22} - g_{23})
 (2 g_{44} + 3 g_{43})))}{64 \pi^2 (18 g_{44}+b')^3 }\beta_{44}\\
&\, +\frac{3 (2 g_{22} - g_{23})}{32 \pi^2(18g_{44}+b')^2}\beta_{43}\,,\\
\beta_{41} =\,&c_4+\frac{7 b'^2 + 252 b' g_{44} + 
 162 (29 g_{44}^2 + 60 g_{44} g_{43} + 60 g_{43}^2)}{8640 \pi^2 (18g_{44}+b')^2}
 -\frac{b' + 27 g_{44} + 18 g_{43}}{96\pi^2 (18g_{44}+b')^2}\beta_{43}\\
 &\,+\frac{b'^2 + 9 b' (9 g_{44} + 10 g_{43}) 
+81(29g_{44}^2+80g_{44}g_{43}+60g_{43}^2)}{960\pi^2(18g_{44}+b')^3}\beta_{44}\,.
\end{split}
\end{equation}

To compare with the beta functions of the previous section,
which were also read off as the coefficients of powers of $R$, we
should identify all the $g_{2i}$'s and all the $g_{4j}$'s above.
We see that these results are clearly very different from the
ones obtained in the Weyl--invariant procedure.
In particular, we observe that $k$ never appears explicitly,
and only the beta function of $g_{41}$ gets a direct
contribution from the matter, via the coefficient $c_4$.

In order to evaluate the beta functions of the couplings $g_{22}$, $g_{23}$, 
$g_{42}$....$g_{46}$, we must now consider the case when $\bar\sigma$ is not constant.
The second variation of \eqref{gen_eff_action} is then
\begin{equation}
\begin{split}
\label{sym_var}
\frac{\delta^2 \Gamma_k^{\rm grav}}{\delta\sigma\delta\sigma}& =
\sqrt{\hat g}
\Bigl[16 g_0 e^{4\bar\sigma}
+4 e^{2\bar\sigma}\left(g_{21}\hat R-6 g_{22}\hat\Box\bar\sigma
-6 g_{23}(\hat\nabla\bar\sigma)^2\right)\\
&-24 g_{23}e^{2\bar\sigma}\hat\nabla^\mu\bar\sigma\hat\nabla_\mu
+24g_{43}\hat\nabla^\mu\hat R\hat\nabla_\mu
-288 g_{45}\big(\hat\Box\bar\sigma\hat\nabla^\mu\bar\sigma\hat\nabla_\mu
+\hat\nabla^\mu(\hat\nabla\bar\sigma)^2\hat\nabla_\mu\big)\\
&+288 g_{46}\hat R^{\mu\nu}\hat\nabla_\mu\bar\sigma\hat\nabla_\nu
+12(g_{23}-2g_{22})e^{2\bar\sigma}\hat\Box
-144 g_{45}\left((\hat\nabla\bar\sigma)^2\hat\Box+
2\hat\nabla^\mu\bar\sigma\hat\nabla^\nu\bar\sigma\hat\nabla_\mu\hat\nabla_\nu\right)\\
&+24 g_{43}\hat R\hat\Box
+288 g_{46}(\hat\nabla^\mu\hat\nabla^\nu\bar\sigma\hat\nabla_\mu\hat\nabla_\nu
-\hat\Box\bar\sigma\hat\Box)
+72g_{44}\hat\Box^2 +4b'\hat\Delta_4
\Bigr]\ .
\end{split}
\end{equation}
Note that this expression is equal to \eqref{S_var} with the couplings $g_i$
appropriately split into $g_{ij}$. 
Since this is no longer a function of $-\hat\Box$ alone, 
we cannot apply a type I cutoff here, as we have done for the constant $\bar\sigma$ case. 
Rather, we shall use a type III procedure, imposing the cutoff on the eigenvalues
of the fourth order operator
\begin{equation}
\hat{\cal O}\equiv \frac{1}{\sqrt{\hat g}(72 g_{44}+4b')}\frac{\delta^2 
\Gamma_k^{\rm grav}}{\delta\sigma\delta\sigma}
=\hat\Box^2+\ldots\,.
\end{equation}
Similarly, given the second variation \eqref{varmatter},
for the local matter contribution
we shall impose the cutoff on the eigenvalues of the 
following operators
\begin{equation}
\begin{split}
\hat{\cal O}_S \equiv & \,\hat\Box + 2\hat\nabla^\mu\bar\sigma\hat\nabla_\mu\ ,
\qquad
\hat{\cal O}_D\equiv \, \hat\Box + 2\hat\nabla^\mu\bar\sigma\hat\nabla_\mu
-\frac{1}{4}\left(\hat R -6\hat\Box\bar\sigma - 6(\hat\nabla\bar\sigma)^2\right)\,,\\
\hat{\cal O}_{M\mu\nu} \equiv &\,\left(\hat\Box
+2\hat\nabla^\lambda\bar\sigma\hat\nabla_\lambda
+\hat\Box\bar\sigma+2(\hat\nabla\bar\sigma)^2\right)\hat g_{\mu\nu}
-R_{\mu\nu}+2\hat\nabla_\mu\hat\nabla_\nu\bar\sigma - 
2\hat\nabla_\mu\bar\sigma\hat\nabla_\nu\bar\sigma\,.
\end{split}
\end{equation}
Limiting ourselves again to a one loop approximation, the FRGE reads
\begin{equation}
\label{simpleII}
\partial_t \Gamma_k = \frac{1}{2}{\rm Tr}
\frac{\partial_t R_k(\hat{\cal O})}{P_k(\hat{\cal O})} 
+\frac{n_S}{2}{\rm Tr}\frac{\partial_t R_k(\hat{\cal O}_S)}{P_k(\hat{\cal O}_S)}
-\frac{n_D}{2}{\rm Tr}\frac{\partial_t R_k(\hat{\cal O}_D)}{P_k(\hat{\cal O}_D)} 
+ \frac{n_M}{2}{\rm Tr}\frac{\partial_t R_k(\hat{\cal O}_M)}{P_k(\hat{\cal O}_M)}
-n_M{\rm Tr}\frac{\partial_t R_k(\hat{\cal O}_S)}{P_k(\hat{\cal O}_S)}\ .
\end{equation}
The heat kernel coefficients that are necessary for the evaluation
of these traces to the desired order are known in the literature, and
we refer to the Appendix for further details on the calculation.
We then obtain the following beta functions:
\begin{equation}
\label{betaIII}
\begin{split}
\beta_{0} =&\frac{1}{32\pi^2}\left(-\frac{8 g_0}{18 g_{44} +b'}
+\frac{9(g_{23}-2g_{22})^2}{(18g_{44}+b')^2}\right)\,,\\
\beta_{21}=&\frac{1}{32\pi^2}\left(
\frac{g_{23}-2g_{21}-2g_{22}}{18 g_{44} +b'} +
\frac{(g_{23}-2g_{22})(36g_{43}-b')}{(18 g_{44} +b')^2}
\right)\,,\\
\beta_{22}=& \frac{1}{32\pi^2}\left(
-\frac{ 2g_{22}}{18 g_{44} +b'}
+\frac{54g_{46}(g_{23}-2g_{22})}{(18 g_{44} +b')^2}
\right)\,,\\
\beta_{23} =&\frac{1}{32\pi^2}\left(-\frac{2 g_{23}}{18 g_{44} +b'} 
- \frac{54g_{45}(g_{23}-2g_{22})}{(18 g_{44} +b')^2}
\right)\,,\\
\beta_{41} =&\,c_4 + \frac{1}{32\pi^2}\left(\frac{29}{540} 
+ \frac{36 g_{43} - b'}{9 (36 g_{44} +2b')}+
\frac{(36g_{43}-b')^2}{9 (36 g_{44} +2b')^2}
\right)\,,\\
\beta_{42}= &\,c'_4+\frac{1}{32\pi^2}\left(
\frac{3 g_{46}}{36 g_{44} +2b'} 
+\frac{6g_{46}(36g_{43}-b')}{(36 g_{44} +2b')^2}\right)
\,,\\
\beta_{43}= &\,c''_4+\frac{1}{32\pi^2}\left(\frac{ 3g_{45}}{36 g_{44} 
+2b'} 
+\frac{6g_{45}(36g_{43}-b') -36 g_{46}^2}
{(36 g_{44} +2b')^2}\right)\,,\\
\beta_{44} =&\, c''_4+
\frac{1}{32\pi^2}\frac{90 g_{46}^2}{(18 g_{44} +b')^2}\,,\\
\beta_{45} =&\, c''_4+
\frac{1}{32\pi^2}\frac{90 g_{45}^2}{(18 g_{44} +b')^2} 
\,,\\
\beta_{46} =&\, c''_4+\frac{1}{32\pi^2}\frac{90 g_{45} g_{46}}{(18 
g_{44} +b')^2}
\,,
\end{split}
\end{equation}
where $c_4$ is defined in \eqref{matcont} and
\begin{equation}
c'_4\equiv \frac{n_S-n_M-n_D}{2304\pi^2}\,,\quad c''_4 \equiv \frac{n_S+2n_M-n_D}{2304\pi^2}\,.
\end{equation}

We first note that the couplings $g_{41}$ and $g_{42}$ do not appear in these
equations, because the corresponding operators contain less than two powers of $\sigma$.
Next, we observe that the beta functions of 
$g_0$, $g_{21}$ and $g_{41}$ are exactly the same as \eqref{betaI} 
at one loop (i.e., neglecting the terms with the $\beta_{ij}$ on the r.h.s.).
As discussed in the previous section, this was fully expected in the case of $g_{41}$.
It is not generally true for the dimensionful couplings
such as $g_0$ and $g_{21}$, but in the present case it is so, 
as all the terms in the beta functions derive from the heat kernel coefficient $B_4$,
whose contributions are scheme independent \cite{cpr1}.
We also note that the second term in $\beta_{41}$ is equal to the
scheme--independent part of the Weyl invariant $\beta_4$, given in \eqref{universal}.
This is another strong indication of the universality of that expression.

We can also explicitly see that the beta functions of the various $g_{2i}$ and $g_{4i}$
are generally not equal, and thus Weyl invariance is broken. 
Even if we started from an initial point where these couplings were the same, 
the flow would lead us away from that situation.
It is remarkable, however, that if we neglect the matter contributions and
set $g_{22}=g_{23}\equiv \hat g_2$ and $g_{44}=g_{45}=g_{46}\equiv \hat g_4$, 
we find that  $\beta_{22} = \beta_{23}\equiv \hat \beta_2$  and 
$\beta_{44}=\beta_{45}=\beta_{46}\equiv \hat\beta_4$. 
In \cite{am1}, the beta functions for conformally reduced gravity in the 
presence of the conformal anomaly were calculated via dimensional
regularization techniques in flat space perturbation theory. If we did
the FRGE calculation above only in flat space, we would not be able
to compute the beta functions of the couplings which multiply operators
containing $\hat R$, namely $g_{21}$, $g_{41}$, $g_{42}$ and $g_{43}$, 
and the remaining beta functions would be exactly the 
$\hat\beta_i$ above, upon equating the couplings. 

In order to compare these with the results in \cite{am1}, we 
make the identifications
\begin{equation}
g_0 = \lambda\ ,
\quad \hat g_2 =-\tfrac{1}{6}\gamma\ , 
\quad \hat g_4=\tfrac{1}{36}\zeta\ ,
\quad Q^2=(4\pi)^2(2b'+\zeta)\ .
\end{equation}
The first three definitions are chosen to agree with the Euclidean 
version of \cite{am1}, which involves a change of sign.
Since the anomaly should be the same independently
of the signature, we do not change the sign of the Riegert action
under Euclidean continuation.
With these definitions, the equations for the couplings $g_{0},\hat g_{2},
\hat g_{4}$ for the flat space case become
\begin{equation}\label{ambeta}
\beta_\lambda = -\frac{8\lambda}{Q^2}+\frac{8 \pi^2\gamma^2}{Q^4}\ ,\quad
\beta_\gamma =  -\frac{2\gamma(Q^2+24\pi^2\zeta)}{Q^4}\ ,\quad
\beta_\zeta = \frac{80\pi^2\zeta^2}{Q^4}\ .
\end{equation}
The equation for $\zeta$ exactly agrees with \cite{am1} in the special
case $\alpha =1$, as does the equation for $\gamma$ when we set
$\zeta=0$, modulo an overall sign.  We find agreement also in the equation 
for $\lambda$ up to non-universal terms, again in the case $\alpha=1$ 
and modulo an overall sign. As we see, it is  the beta functions 
in the Weyl-breaking procedure that reproduce the results of \cite{am1}.
In fact, one could assume that $\sigma$ scales anomalously under \eqref{weyl}
as $\sigma\to\sigma-\alpha\omega$ and one would then also recover the
$\alpha$--dependence discussed in \cite{am1}. We will not discuss
this here.

However, as it turns out, this procedure breaks not only Weyl invariance,
but also global scale invariance. That is to say, in addition to the ratios between
the coefficients of the operators in \eqref{eff_action} being different, as we
have seen from the full set of beta functions above, new terms not 
originally present in the action are also generated, and hence not even the
form \eqref{gen_eff_action} is preserved. 
These new terms will contribute to the beta functions \eqref{betaIII}
and will themselves have non-zero beta functions.
For example, expanding the trace in  \eqref{breaking}, the matter 
contributions proportional to $c_0$ and $c_2$ multiply the operators 
$\int dx\sqrt{\hat  g}$ and $\int dx\sqrt{\hat  g}\hat R$, and 
we will also have operators such as $\int dx\sqrt{\hat  g}\,e^{2\bar\sigma}$,
$\int dx\sqrt{\hat  g}\,e^{4\bar\sigma}\hat R$, etc.
The flow thus takes place in a much larger class of actions,
where the dependence on $\hat g_{\mu\nu}$ and $\sigma$
is not restricted by the demand of invariance under \eqref{weyl}.

Nonetheless, we do not expect these new terms to contribute to
the beta functions of the $g_{4i}$ above, as the new terms will come with powers 
of $e^{\bar\sigma}$ which do not correspond to those of the operators multiplying 
the couplings $g_{41}\ldots g_{46}$ in \eqref{gen_eff_action}. 
For the same reason, we do not expect the couplings in \eqref{gen_eff_action} to be present in 
the new beta functions $\beta_{4j}$  ($j > 6$), apart from $g_{44}$ contributions
in the denominator. Thus, we can already 
say something on the existence of fixed points by considering the $\beta_{4i}$ 
that we have written.

From \eqref{ambeta}, we note that the beta function for $\zeta$ vanishes
in the case $\zeta=0$, in accordance with \cite{am1}. In terms of 
the couplings $g_{44}$, $g_{45}$ and $g_{46}$ this is equivalent to the 
vanishing of those beta functions for $g_{45}=g_{46}=0$, neglecting the local matter
contribution. Remarkably, the beta functions for $g_{42}$ and $g_{43}$ also
vanish in this situation. 
The remaining beta function for $g_{41}$, on the other hand, is non-vanishing 
for any real value of the couplings, and one might be tempted to conclude that
there is then no FP solution. But $g_{41}$ is not a coupling for conformally reduced 
gravity in this Weyl breaking setting, since the corresponding operator does not contain 
the dynamical  field $\sigma$ and, unlike in the Weyl invariant case,
$g_{41}$ is independent of the other fourth-order couplings.
Therefore, there is no reason to require that its beta function vanish.
If we do not impose the vanishing of $\beta_{41}$, we find agreement with 
the results of \cite{am1}, at least within the restricted set of beta functions 
\eqref{ambeta}.
In order to draw more general conclusions, however, one would have to
study the flow of the other couplings that have not been included
in the action \eqref{gen_eff_action}, but which will be generated
by quantum effects.

\section{Conclusions}

We have reconsidered the calculation of beta functions
in conformally reduced gravity.
Our main tool has been the functional renormalization group equation, 
which is well suited to discuss the RG flow of nonrenormalizable
theories, and has been used very effectively to find a gravitational
fixed point in various approximations.
In the present context, the dynamics of gravity has been
essentially reduced to that of a scalar field.
From a physical point of view, it is not clear that this severe
truncation still captures the essential features of gravity.
It has been argued in \cite{am1,amm1} that it does so in the
extreme infrared, and as shown in \cite{reuterweyerI, reuterweyerII}
in the Einstein-Hilbert truncation
it also yields a fixed point in the UV with properties that are quite close
to those obtained in the presence of the transverse gravitons.
From a theoretical point of view, it has the advantage that 
it sidesteps several issues, such as gauge fixing,
which do arise in the complete formulation of gravity.
Thus, the conformal reduction may be at least a good theoretical laboratory
in which to test various ideas.

In this paper we have considered, in addition to local terms up to second order
in curvature, also the nonlocal terms responsible for the conformal anomaly
of massless matter fields.
This nonlocal action depends only on the number of massless fields
and is thus not itself subject to RG flow, in agreement with the findings of
 \cite{machadosaueressig}. It does, however, affect the running of the other
couplings.

Following the general discussion in \cite{floreanini,reuterweyerI},
the IR cutoff can be implemented in two inequivalent ways,
which either maintain or break the Weyl invariance (\ref{weyl}).
They are both mathematically consistent procedures.
In fact we have already observed in the end of Section 1 that the Weyl
transformations \eqref{weyl} should not be regarded as a gauge invariance
in conformally reduced gravity, and therefore it is not mandatory to preserve them
in the quantum theory.
From the physical standpoint, one could try to interpret this choice
as that between treating the cutoff scale $k$ as internal to the theory
(when $\bar g_{\mu\nu}$ is used to define the cutoff), 
or as an absolute external scale (when $\hat g_{\mu\nu}$ is used to define the cutoff).
Which of the two procedures correctly describes quantum gravity is
something that, in our opinion, can only be assessed by observation or experiment.
It is tempting to speculate, nevertheless, that the correct procedure to be
used in the description of UV physics is the first one.
We note that at low energies there are various sets of phenomena that
define dynamical mass scales:
electroweak physics determines the mass of the electron and hence
atomic spectroscopy,
strong interaction physics determines the mass of the nucleons.
Both of these scales are to a large extent unaffected by gravity,
and in principle one could use electroweak or strong mass units
to define the fiducial metric $\hat g$ that is used in the second type of cutoff
\footnote{On a historical note, it was precisely the availability
of these absolute units that formed the basis of Einstein's critique
of Weyl's theory.}.
When one considers very high energy phenomena, however, such as the universe at the
GUT energy scale, neither atoms nor nuclei, nor even the VEV of the
Higgs field, are there to provide an absolute reference scale,
and in any case gravity is then so strong that its influence
cannot be neglected. In such circumstances it seems that only
the former procedure is meaningful.

In the case of the Weyl invariant procedure, 
we have extended the results of \cite{reuterweyerI} 
by including the contributions of matter, and the curvature squared term. 
It appears that a physically acceptable fixed point is not present in this truncation
in pure gravity, but that it reappears in the presence of suitable matter fields.
It may or may not reappear in pure conformally reduced gravity when higher order terms are 
included.
We should also mention that a fixed point with the
correct properties does not appear if we 
restrict ourselves to conformal fluctuations {\it after}
having expanded the action.
This is somehow to be expected, since scalar fields tend to
generate a fixed point with negative $G$.
In any case, it is worth stressing that this negative result does not have 
direct implications for the asymptotic safety programme.

We have then shown that the Weyl breaking procedure leads to 
beta functions which are very different from the invariant ones, but which 
generally agree with those given in \cite{am1}, at least as far as the 
case of a flat space background is concerned.
This RG flow, however, will break not only Weyl invariance, but also
global scale invariance, and it will hence generate new couplings that are not
present in the class of actions that we have considered. As these new couplings
will have non-zero beta functions, a proper 
discussion of the fixed points in this theory would require an extension of our 
current analysis, which we leave for future work.

\medskip

\subsection*{Acknowledgements}
We would like to thank R. Loll, D. Litim and O. Rosten for discussions,
and I. Antoniadis and E. Mottola for stimulating correspondence.
R.P. also wishes to thank R. Loll and D. Litim for hospitality at the University
of Utrecht and at the University of Sussex, where this work was respectively 
initiated and concluded.
R.P. is supported in part by a grant of the Royal Society. P.F.M. is supported by the 
Netherlands Organization for Scientific Research (NWO) under their VICI program.

\medskip

\begin{appendix}
\section{Trace evaluation for type III cutoffs}

In this Appendix, we collect some of the formulas necessary for the evaluation
of the operator traces in Sections 4 and 5 (for more details, see, e.g., Appendix
A in \cite{cpr1}). Generally,
the traces of the functions $W(\Delta)$ appearing on the r.h.s. of the FRGE may 
be evaluated via the asymptotic keat kernel expansion
\begin{equation}\label{HKE}
Tr W(\Delta) = \frac{1}{(4\pi)^{2}}\Bigl[Q_{\frac{4}{p}}(W){\rm B}_0(\Delta)+
Q_{\frac{2}{p}}(W){\rm B}_2(\Delta)+ Q_{0}(W){\rm B}_d(\Delta)+\ldots\Bigr]\,,
\end{equation}
where  $\Delta$ is an elliptic operator of order $p$
and 
\begin{equation}
Q_n(W) = \frac{1}{\Gamma(n)} \int_0^\infty dz z^{n-1} W(z),
\end{equation}
for $n>0$, while $Q_0(W)=W(0)$.
For the specific cases of the operators $\hat{\cal O}_i$ appearing in 
\eqref{simpleII}, the heat kernel coefficients 
may be computed using the formulas in, e.g., \cite{barth,lps}, reading
\begin{equation}
\begin{split}
{\rm B}_0(\hat{\cal O}) = & \int dx\,\sqrt{\hat g} \,,\\
{\rm B}_2(\hat{\cal O}) =& \sqrt{\pi}\int dx\,\sqrt{\hat g} \bigg\{
\left[\frac{1}{12} +\frac{36 g_{43} -b'}{12\tilde Q}\right] \hat R 
+\frac{3(g_{23} - 2 g_{22})}{2\tilde Q} e^{2\bar\sigma}
-\frac{27  g_{45} }{\tilde Q} (\hat\nabla\bar\sigma)^2 
-\frac{ 27   g_{46}}{\tilde Q} \hat\Box\bar\sigma
\bigg\}\,,\\
{\rm B}_4(\hat{\cal O})=& \int dx\,\sqrt{\hat g} \bigg\{
\left[-\frac{4 g_0}{\tilde Q}
+\frac{9(g_{23}-2g_{22})^2}{\tilde Q^2}\right] e^{4\bar\sigma}\\
&
+\left[ \frac{g_{23}-2g_{22}-2g_{21}}{2\tilde Q} 
+\frac{(g_{23}-2g_{22})(36g_{43}-b')}{\tilde Q^2}
\right]e^{2\bar\sigma}\hat R\\
& +\left[ \frac{29}{2160} + \frac{36g_{43}-b'}{36 \tilde Q}+
\frac{(36g_{43}-b')^2}{36 \tilde Q^2}
\right] \hat R^2\\
&+\left[\frac{6 g_{22}}{\tilde Q}
-\frac{324 g_{46}(g_{23}-2g_{22})}{\tilde Q^2}
\right]   e^{2\bar\sigma}\hat\Box\bar\sigma
+\left[\frac{6 g_{23}}{\tilde Q}
-\frac{324g_{45}(g_{23}-2g_{22})}{\tilde Q^2}
\right]   e^{2\bar\sigma}(\hat\nabla\bar\sigma)^2\\
&- \left[\frac{9 g_{46}}{\tilde Q}
+\frac{18g_{46}(36g_{43}-b')}{\tilde Q^2}
\right]   \hat R \hat \Box\bar\sigma
-\left[\frac{9 g_{45}}{\tilde Q}
+\frac{18g_{45}(36g_{43}-b')+108 g_{46}^2}
{\tilde Q^2}\right]   \hat R (\hat\nabla\bar\sigma)^2\\
&+\frac{6480   g_{45} g_{46}}{\tilde Q^2}(\hat\nabla\bar\sigma)^2 \hat \Box\bar\sigma
+\frac{3240g_{45}^2}{\tilde Q^2} ((\hat\nabla\bar\sigma)^2)^2
+\frac{3240 g_{46}^2}{\tilde Q^2}(\bar \Box\bar\sigma)^2\bigg\}\,,
\end{split}
\end{equation}
with $\tilde Q \equiv (36 g_{44} +2b')$, and
\begin{equation}
\begin{split}
{\rm B}_0^{\rm mat} = & \int d^4x\,\sqrt{\hat g} 32\pi^2 c_0\,,\\
{\rm B}_2^{\rm mat} = & \int d^4x\,\sqrt{\hat g} \Bigl\{32\pi^2 c_2 \hat R
+ (n_S+8n_M-10n_D) \hat\Box\bar\sigma -(n_S-4n_M+2n_D)(\hat\nabla\bar\sigma)^2\Bigr\}\,,\\
{\rm B}_4^{\rm mat} = & \int d^4x\,\sqrt{\hat g} \Bigl\{32\pi^2 c_4 \hat R^2
+ \frac{n_S +2 n_M-n_D}{2}(\hat\Box\bar\sigma)^2 
+  \frac{n_S +2n_M-n_D}{2}((\hat\nabla\bar\sigma)^2)^2\\
&\quad- \frac{n_S -n_M-n_D}{6}\hat R \hat\Box\bar\sigma
-\frac{n_S +2n_M-n_D}{6}\hat R (\hat\nabla\bar\sigma)^2
+(n_S+2 n_M-n_D)\Box\bar\sigma (\hat\nabla\bar\sigma)^2
 \Bigr\}\,,
\end{split}
\end{equation}
where for convenience we have collected the coefficients from the
local matter contribution under the ${\rm B}^{\rm mat}_i$. 
The case of the Weyl-invariant operators appearing in \eqref{simple} in
Section 4 may be readily obtained from the
above by letting $\hat g_{\mu\nu} \mapsto \bar g_{\mu\nu}$ and 
$\bar\sigma \mapsto 0$. 

Using the generalized optimized cutoff $R_k(z) = (k^p - z)\Theta(k^p-z)$
for the $pth$-order operators, the functions $Q_i\left(\tfrac{\partial_t R_k}{P_k}\right)$ 
may also be straightforwardly evaluated. Imposing 
the cutoff on the $\hat{\cal O}_i$ from Sections 4 and 5,
we find
\begin{equation}
Q_0 = p\,,\qquad
Q_{\frac{1}{2}}=\frac{2p}{\sqrt{\pi}}k^{p/2}\,,\qquad
Q_1 = p k^p\,,\qquad
Q_2 = \frac{p}{2} k^{2p}\,.
\end{equation}

\end{appendix}


\begin{thebibliography}{99}

\bibitem{ghp}
G.W. Gibbons, S.W. Hawking and M.J. Perry,
Nucl. Phys. {\bf B138} 141 (1978)barth

\bibitem{mm1}
P.O. Mazur and E. Mottola, Nucl. Phys. {\bf B341} 187-212 (1990);

\bibitem{mottolameasure}
E. Mottola, J. Math. Phys. {\bf 36} 2470-2511 (1995). 
e-Print: hep-th/9502109

\bibitem{dasguptaloll}
A. Dasgupta and R. Loll, Nucl. Phys. {\bf B 606}357-379 (2001) 
e-Print: hep-th/0103186

\bibitem{ajl}
J.~Ambj\o rn, A.~Dasgupta, J.~Jurkiewicz and R.~Loll,
Nucl.\ Phys.\ Proc.\ Suppl.\  {\bf 106}, 977 (2002)
e-Print:hep-th/0201104
J.~Ambj\o rn, J.~Jurkiewicz and R.~Loll,
Phys.\ Rev.\  D {\bf 72}, 064014 (2005)
e-Print:hep-th/0505154
J.~Ambj\o rn, A.~Gorlich, J.~Jurkiewicz and R.~Loll,
Phys.\ Rev.\  D {\bf 78}, 063544 (2008)
e-Print:arXiv:0807.4481 [hep-th]

\bibitem{riegert}
R.J. Riegert, Phys. Lett. {\bf B 134} 56-60 (1984).

\bibitem{am1}
I. Antoniadis and E. Mottola, Phys. Rev. {\bf D45} 2013 (1992)

\bibitem{amm1}
I. Antoniadis, P.O. Mazur and E. Mottola, Nucl. Phys. {\bf B 388} 627-647 (1992)
e-Print: hep-th/9205015
Phys. Lett. {\bf B323} 284-291 (1994) 
e-Print: hep-th/9301002
Phys. Rev. {\bf D55} 4756-4769 (1997)
e-Print: hep-th/9509168
Phys. Rev. {\bf D55} 4770-4784 (1997)
e-Print: hep-th/9509169
Phys. Lett. {\bf B394} 49-56 (1997) 
e-Print: hep-th/9611145

\bibitem{mazurmottola}
P.O.  Mazur and E. Mottola, Phys. Rev. {\bf D 64} 104022 (2001) 
e-Print: hep-th/0106151


\bibitem{amm2}
I. Antoniadis, P.O.  Mazur and E. Mottola, Phys. Rev. Lett. {\bf 79} 14-17 (1997) 
e-Print: astro-ph/9611208
Phys. Lett. {\bf B 444} 284-292 (1998) 
e-Print: hep-th/9808070
New J. Phys. {\bf 9} 11 (2007) 
e-Print: gr-qc/0612068

\bibitem{mottolavaulin}
E. Mottola and R. Vaulin, Phys. Rev. {\bf D 74} 064004 (2006) 
e-Print: gr-qc/0604051


\bibitem{reuter}
M. Reuter, Phys. Rev. {\bf D57}, 971 (1998) [arXiv:hep-th/9605030].

\bibitem{dou}
D. Dou and R. Percacci, Class. Quant. Grav. {\bf 15} 3449 (1998);
[arXiv:hep-th/9707239].

\bibitem{souma} 
W. Souma, Prog. Theor. Phys. {\bf 102}, 181 (1999); [arXiv:hep-th/9907027].

\bibitem{reuterEH} O. Lauscher and M. Reuter, Phys. Rev. {\bf D65}, 025013 (2002);
[arXiv:hep-th/0108040];
Class. Quant. Grav. {\bf 19}, 483 (2002);
[arXiv:hep-th/0110021];
Int. J. Mod. Phys. {\bf A 17}, 993 (2002);
[arXiv:hep-th/0112089];
M. Reuter and F. Saueressig, Phys. Rev. {\bf D65}, 065016 (2002).
[arXiv:hep-th/0110054].

\bibitem{fischerlitim}
D. F. Litim, 
Phys. Rev. Lett. {\bf 92} 201301 (2004);
[arXiv:hep-th/0312114];
AIP Conf.\ Proc.\  {\bf 841} (2006) 322
[arXiv:hep-th/0606044];
e-Print: arXiv:0810.3675 [hep-th], 
To appear in the proceedings of 
"From Quantum to Emergent Gravity: Theory and Phenomenology", 
June 11-15 2007, Trieste, Italy;
P. Fischer and D. F. Litim,
AIP Conf.\ Proc.\  {\bf 861} (2006) 336
[arXiv:hep-th/0606135];
Phys. Lett. {\bf B 638} (2006) 497
[arXiv:hep-th/0602203]


\bibitem{lauscherreuter}
O. Lauscher and M. Reuter, Phys. Rev. {\bf D 66}, 025026 (2002)     
[arXiv:hep-th/0205062].

\bibitem{codellopercacci}
A. Codello and R. Percacci, Phys. Rev. Lett. {\bf 97}, 221301 (2006);
e-Print: hep-th/0607128.

\bibitem{Benedetti:2009rx}
  D.~Benedetti, P.~F.~Machado and F.~Saueressig,
  arXiv:0901.2984 [hep-th];
  arXiv:0902.4630 [hep-th].

\bibitem{largen}
R. Percacci,
Phys. Rev. {\bf D73}, 041501(R) (2006);
[arXiv:hep-th/0511177].

\bibitem{cpr1}
A. Codello, R. Percacci and C. Rahmede,
Int.\ J.\ Mod.\ Phys.\  A {\bf 23} 143 (2008);
[arXiv:0705.1769 [hep-th]].
Ann. Phys. {\bf 324}, 414-469 (2009),
e-Print: arXiv:0805.2909 [hep-th]

\bibitem{machadosaueressig}
  P.~F.~Machado and F.~Saueressig,
  arXiv:0712.0445 [hep-th].



\bibitem{rs}
M. Reuter and F. Saueressig,
Phys. Rev. {\bf D 66} 125001 (2002)
e-Print: hep-th/0206145.

\bibitem{niedermaierreuter}
M. Niedermaier and M. Reuter,
Living Rev. Relativity 9,  (2006), 5.

\bibitem{niedermaierrev} 
M. Niedermaier,
Class. Quant. Grav.  {\bf 24} (2007) R171
[arXiv:gr-qc/0610018].

\bibitem{revperc}
R. Percacci, 
``Asymptotic Safety'',
to appear in ``Approaches to Quantum Gravity: Towards a New Understanding of Space, Time and Matter''
ed. D. Oriti, Cambridge University Press;
e-Print: arXiv:0709.3851 [hep-th].

\bibitem{rw1} 
M. Reuter and F. Saueressig,
Fortsch. Phys. {\bf 52} 650-654 (2004)
e-Print: hep-th/0311056;
M. Reuter, H. Weyer, 
Phys. Rev. {\bf D 70} 124028 (2004) 
e-Print: hep-th/0410117;
JCAP 0412:001 (2004) 
e-Print: hep-th/0410119;
Int. J. Mod. Phys. {\bf D15} 2011-2028 (2006) 
e-Print: hep-th/0702051,

\bibitem{reutercosmology}
  M.~Reuter and F.~Saueressig,
  JCAP {\bf 0509} (2005) 012
  [arXiv:hep-th/0507167];
  A.~Bonanno and M.~Reuter,
  JCAP {\bf 0708} (2007) 024
  [arXiv:0706.0174 [hep-th]].

\bibitem{niedermaier}
M. Niedermaier, JHEP 0212:066 (2002), 
e-Print: hep-th/0207143.
M. Niedermaier, Nucl. Phys. {\bf B 673} 131-169 (2003) 
e-Print: hep-th/0304117

\bibitem{reuterweyerI}
M. Reuter, H. Weyer, MZ-TH-08-04, e-Print: arXiv:0801.3287 [hep-th]
\bibitem{reuterweyerII}
M. Reuter, H. Weyer, MZ-TH-08-12, e-Print: arXiv:0804.1475 [hep-th]

\bibitem{duff}
M.J. Duff, Nucl. Phys. {\bf B 125} 334 (1977).

\bibitem{wetterich} 
C. Wetterich, Phys. Lett. {\bf B 301} 90-94 (1993).


\bibitem{floreanini} 
R. Floreanini, R. Percacci, Nucl. Phys. {\bf B436} 141-162 (1995)
e-Print: hep-th/9305172

\bibitem{bbm}
Z. Bern, E. Mottola and S.K. Blau
Phys. Rev. {\bf D 43} 1212-1222 (1991).

\bibitem{manriquereuter}
E. Manrique and M. Reuter, e-Print: arXiv:0811.3888 [hep-th]

\bibitem{reuterwetterich}
M. Reuter, C. Wetterich, Nucl. Phys. {\bf B 506} 483-520 (1997)

\bibitem{optimized}
D. Litim, Phys.Rev. {\bf D 64} 105007 (2001), e-Print: hep-th/0103195
barth
\bibitem{perini}
R. Percacci and D. Perini, Phys. Rev. {\bf D 67} 081503 (2003) 
e-Print: hep-th/0207033
{\it ibid} 044018 (2003), 
e-Print: hep-th/0304222

\bibitem{percacci}
R. Percacci, J. Phys. {\bf A 40}4895-4914 (2007) e-Print: hep-th/0409199

\bibitem{barth}
  N.~H.~Barth,
  J.\ Phys.\ A  {\bf 20}, 857 (1987);
  J.\ Phys.\ A  {\bf 20}, 875 (1987).
  
\bibitem{lps}
  H.~W.~Lee, P.~Y.~Pac and H.~K.~Shin,
  Phys.\ Rev.\  D {\bf 35}, 2440 (1987).

\end{thebibliography}
\end{document}